\PassOptionsToPackage{unicode}{hyperref}
\PassOptionsToPackage{hyphens}{url}
\PassOptionsToPackage{dvipsnames,svgnames,x11names}{xcolor}
\documentclass[
  authoryear,
  preprint,
  5p,
  twocolumn]{elsarticle}

\usepackage{amsmath,amssymb}
\usepackage{iftex}
\ifPDFTeX
  \usepackage[T1]{fontenc}
  \usepackage[utf8]{inputenc}
  \usepackage{textcomp} 
\else 
  \usepackage{unicode-math}
  \defaultfontfeatures{Scale=MatchLowercase}
  \defaultfontfeatures[\rmfamily]{Ligatures=TeX,Scale=1}
\fi
\usepackage{lmodern}
\ifPDFTeX\else  
\fi
\IfFileExists{upquote.sty}{\usepackage{upquote}}{}
\IfFileExists{microtype.sty}{
  \usepackage[]{microtype}
  \UseMicrotypeSet[protrusion]{basicmath} 
}{}
\makeatletter
\@ifundefined{KOMAClassName}{
  \IfFileExists{parskip.sty}{%
    \usepackage{parskip}
  }{
    \setlength{\parindent}{0pt}
    \setlength{\parskip}{6pt plus 2pt minus 1pt}}
}{
  \KOMAoptions{parskip=half}}
\makeatother
\usepackage{xcolor}
\ifx\paragraph\undefined\else
  \let\oldparagraph\paragraph
  \renewcommand{\paragraph}[1]{\oldparagraph{#1}\mbox{}}
\fi
\ifx\subparagraph\undefined\else
  \let\oldsubparagraph\subparagraph
  \renewcommand{\subparagraph}[1]{\oldsubparagraph{#1}\mbox{}}
\fi

\providecommand{\tightlist}{%
  \setlength{\itemsep}{0pt}\setlength{\parskip}{0pt}}\usepackage{longtable,booktabs,array}
\usepackage{calc} 
\usepackage{etoolbox}
\makeatletter
\patchcmd\longtable{\par}{\if@noskipsec\mbox{}\fi\par}{}{}
\makeatother
\IfFileExists{footnotehyper.sty}{\usepackage{footnotehyper}}{\usepackage{footnote}}
\makesavenoteenv{longtable}
\usepackage{graphicx}
\makeatletter
\def\maxwidth{\ifdim\Gin@nat@width>\linewidth\linewidth\else\Gin@nat@width\fi}
\def\maxheight{\ifdim\Gin@nat@height>\textheight\textheight\else\Gin@nat@height\fi}
\makeatother
\setkeys{Gin}{width=\maxwidth,height=\maxheight,keepaspectratio}
\makeatletter
\def\fps@figure{htbp}
\makeatother

\usepackage{fancyhdr}
\makeatletter
\@ifpackageloaded{caption}{}{\usepackage{caption}}
\AtBeginDocument{%
\ifdefined\contentsname
  \renewcommand*\contentsname{Table of contents}
\else
  \newcommand\contentsname{Table of contents}
\fi
\ifdefined\listfigurename
  \renewcommand*\listfigurename{List of Figures}
\else
  \newcommand\listfigurename{List of Figures}
\fi
\ifdefined\listtablename
  \renewcommand*\listtablename{List of Tables}
\else
  \newcommand\listtablename{List of Tables}
\fi
\ifdefined\figurename
  \renewcommand*\figurename{Figure}
\else
  \newcommand\figurename{Figure}
\fi
\ifdefined\tablename
  \renewcommand*\tablename{Table}
\else
  \newcommand\tablename{Table}
\fi
}
\@ifpackageloaded{float}{}{\usepackage{float}}
\floatstyle{ruled}
\@ifundefined{c@chapter}{\newfloat{codelisting}{h}{lop}}{\newfloat{codelisting}{h}{lop}[chapter]}
\floatname{codelisting}{Listing}

\makeatother
\makeatletter
\@ifpackageloaded{caption}{}{\usepackage{caption}}
\@ifpackageloaded{subcaption}{}{\usepackage{subcaption}}
\makeatother
\usepackage{float}
\makeatletter
\let\oldlt\longtable
\let\endoldlt\endlongtable
\def\longtable{\@ifnextchar[\longtable@i \longtable@ii}
\def\longtable@i[#1]{\begin{figure}[H]
\onecolumn
\begin{minipage}{0.5\textwidth}
\oldlt[#1]
}
\def\longtable@ii{\begin{figure}[H]
\onecolumn
\begin{minipage}{0.5\textwidth}
\oldlt
}
\def\endlongtable{\endoldlt
\end{minipage}
\twocolumn
\end{figure}}
\makeatother
\journal{ASA Student Paper Competition}
\ifLuaTeX
  \usepackage{selnolig}  
\fi
\usepackage[]{natbib}
\usepackage{bookmark}

\IfFileExists{xurl.sty}{\usepackage{xurl}}{} 
\hypersetup{
  pdftitle={A reproducible pipeline for extracting representative signals from wire cuts},
  pdfauthor={Yuhang Lin; Heike Hofmann},
  pdfkeywords={data visualization, striation marks, cross-correlation
function, Hough transformation, forensic statistics},
  colorlinks=true,
  linkcolor={blue},
  filecolor={Maroon},
  citecolor={Blue},
  urlcolor={Blue},
  pdfcreator={LaTeX via pandoc}}

\begin{document}

\begin{frontmatter}
\title{A reproducible pipeline for extracting representative signals
from wire cuts}
\author[1]{Yuhang Lin%
\corref{cor1}%
}
 \ead{yhlin@iastate.edu} 
\author[1]{Heike Hofmann%
}
 \ead{hofmann@iastate.edu} 

\affiliation[1]{organization={Iowa State University, Department of
Statistics},addressline={2438 Osborn
Dr},city={Ames},postcode={50010},postcodesep={}}

\cortext[cor1]{Corresponding author}

\begin{abstract}
We propose a reproducible pipeline for extracting representative signals
from 2D topographic scans of the tips of cut wires. The process fully
addresses many potential problems in the quality of wire cuts, including
edge effects, extreme values, trends, missing values, angles, and
warping. The resulting signals can be further used in source
determination, which plays an important role in forensic examinations.
With commonly used measurements such as the cross-correlation function,
the procedure controls the false positive rate and false negative rate
to the desirable values as the manual extraction pipeline but
outperforms it with robustness and objectiveness.
\end{abstract}

\end{frontmatter}
    
\clearpage
\newpage

\setcounter{page}{1}
\pagestyle{fancy}
\fancyhead{} 
\fancyfoot{} 
\fancyhead[L]{\sc{A reproducible pipeline for extracting representative signals from wire cuts}}
\fancyhead[R]{Y. Lin}
\fancyfoot[R]{\thepage}

\section{Introduction}\label{sec-Introduction}

Determining the source of evidence is a crucial aspect of forensic
examinations. During an investigation, the question of the source might
turn into the specific source problem: Did this tool leave the marks
visible on the evidence? Current practice in forensic labs is that the
examiner will reproduce the crime scene evidence using matching
materials (i.e., fire a bullet of the same type of ammunition from a
suspect's gun or cut the same material with a suspect's tool) and
compare marks on the resulting piece under a comparison microscope to
the marks on the evidence. Examiners summarise the result of these
comparisons according to the AFTE Theory of identification \citep{afte}
as an `identification' (i.e., the suspect's tool made the mark to the
exclusion of any other tool in existence), an `exclusion' (i.e., the
crime scene evidence was made by a different tool), or an `inconclusive'
(i.e., cannot tell either way) result. This practice was first
criticized by the National Research Council \citep{nas2009} and later by
the President's Council of Advisors \citep{pcast} for its subjectivity
and lack of established error rates.

In response to this criticism, the research community has invested in
collecting and distributing validated data
\citep{maNISTBulletSignature2004, zhengNISTBallisticsToolmark2016},
introduced quantitative measures for evaluating the similarity of marks
\citep{chuAutomaticIdentificationBullet2013, vorburgerApplicationsCrosscorrelationFunctions2011},
and suggested automatic matching algorithms
\citep{hareAutomaticMatchingBullet2017, taiFullyAutomaticMethod2018, baiker-sorensenAutomatedInterpretationComparison2023}.

Most of this work was developed with a focus on firearms evidence and
centers on comparisons of breech face impressions on fired cartridge
cases and striation marks on bullets. However, it has been shown
\citep{cuellarRevolutionizingForensicToolmark2023, krishnanAdaptingChumbleyScore2019}
that some of the algorithms depend mostly on the type of marks made and
can be applied more generally.

\begin{figure}

\centering{

\includegraphics[width=0.465\textwidth,height=\textheight]{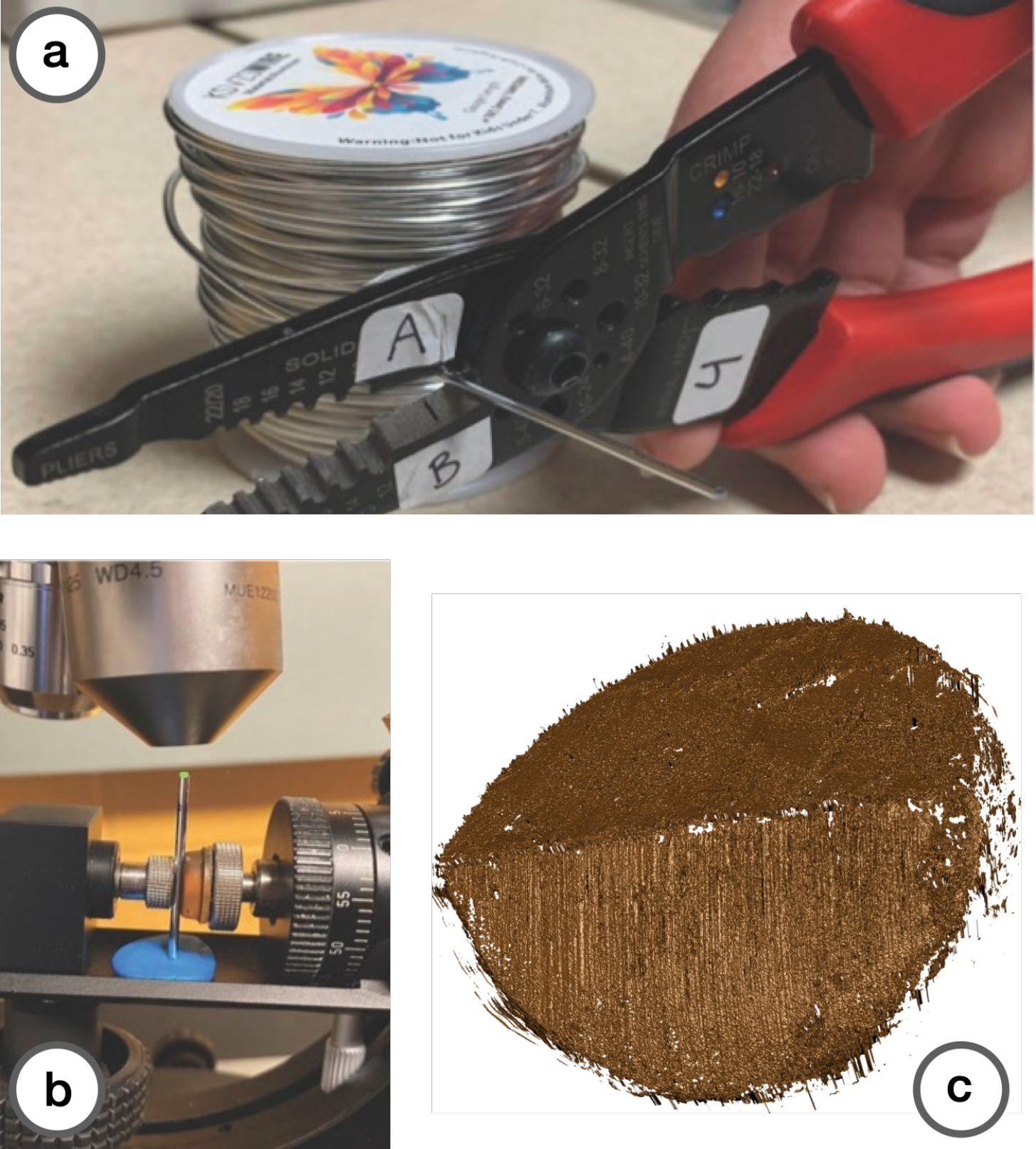}

}

\caption{\label{fig-overview}(a) Wirecutter with labeled blades cutting
2 mm aluminum wire. (b) The cut wire surface is scanned using a confocal
microscope. (c) Rendering of the scanned surface reveals a tent
structure.}

\end{figure}

Here, we focus on the process of extracting signals from 3D topographic
scans of cut wires to assess the similarity of the striation marks
engraved onto the wire surface during the cutting process (see
Figure~\ref{fig-overview}).

\begin{figure}

\centering{

\includegraphics{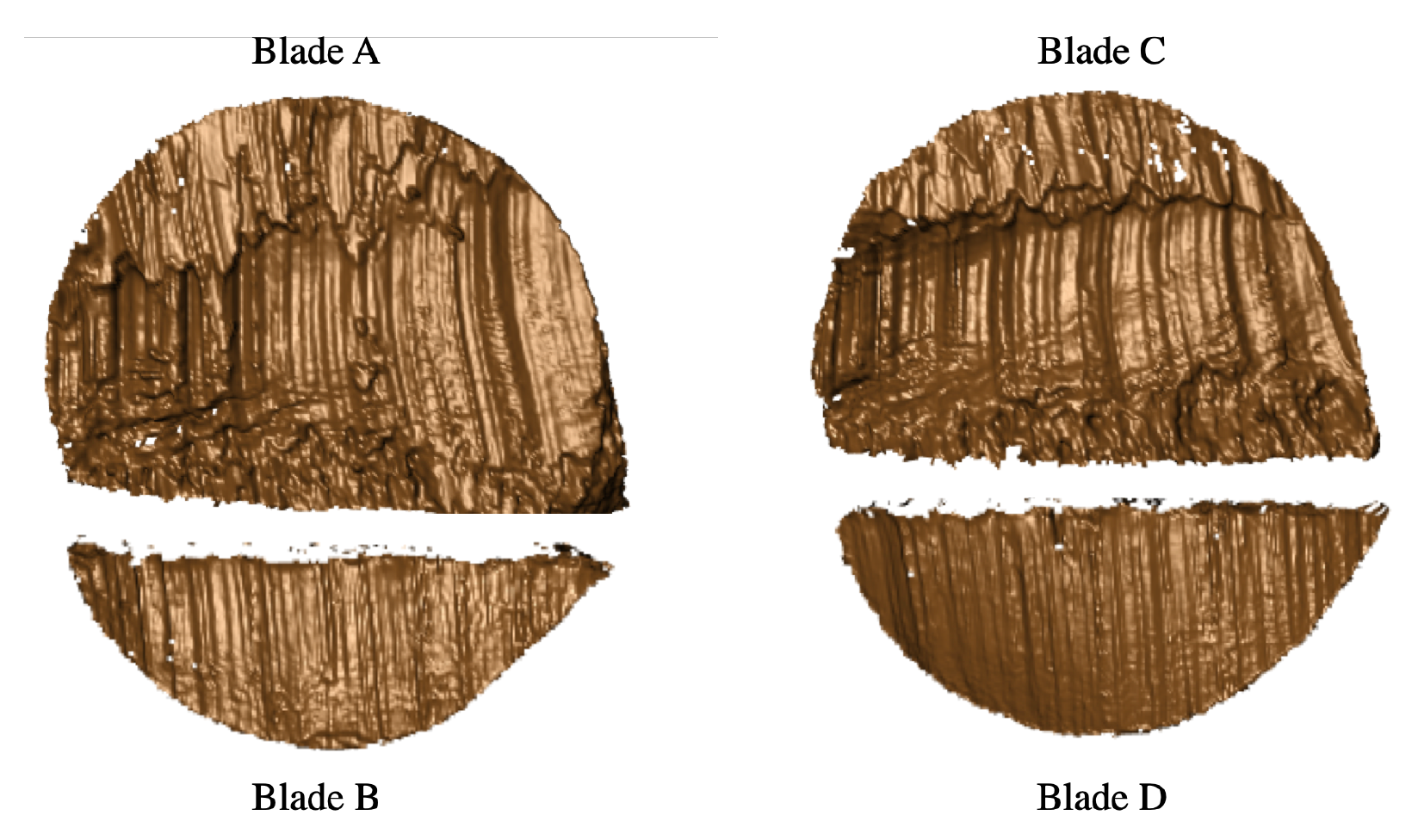}

}

\caption{\label{fig-T1W}Scans of cut wires (tool 1)}

\end{figure}

When cutting a wire with a bladed tool, small imperfections on the
blades touch the wire, scratch over the surface, and leave marks. These
marks appear in the form of striations, as visible in
Figure~\ref{fig-overview}(c). We can think of the striation marks as
repeated observations of the same 2D signal orthogonal to the cutting
direction. The goal, therefore, is to find a curve representative of the
signal engraved on the wire by the tool during the cutting process. Once
we have determined a representative signal for a scan, we can use one of
the procedures in \citet{hareAutomaticMatchingBullet2017},
\citet{chumbleyValidationToolMark2010}, or
\citet{juJournalOpenSourceImplementation2022} to quantify the similarity
between scans of cuts made by the same tool and cuts made by different
tools. Identifying a representative signal in wire scans is more
challenging than in other scans, such as screwdriver marks or bullets,
because of the absence of macro structures (such as shoulders on either
end of the engraved area), which could be used for aligning objects
during the scanning process. Here, wire cuts are aligned under the
microscope along the micro-feature of the ridge of the tent shown in
Figure~\ref{fig-overview}(c) such that each side of the tent's roof is
as parallel to the microscope's lens as possible, creating two
dome-shaped scans for each cut. Figure~\ref{fig-T1W} shows all four
scans corresponding to the two wire ends cut with tool 1 (in the
location closest to the jaw). Each scan shows markings engraved by one
of the blades.

Each scan consists of height measurements taken on a regular grid of
0.645 \(\times\) 0.645 square microns (1 micron = 1 \(\mu\)m = 1/1000
millimeter). The surface measurements can be stored in the form of a
real-valued matrix \(F\) with dimensions height \(h\) \(\times\) width
\(w\). We will refer to an element in this matrix as \(F_{ij}\) with
\(1 \le i \le h\) and \(1 \le j \le w\). We can think of this matrix as
a set of functions across the width of the scan. We will denote the
function corresponding to the \(i\)th row of values in \(F\) as \(f_i\).
The goal of finding a representative signal is equivalent to identifying
a function \(\tilde{f}\) that is representative of the functions
\(f_i\), \(1 \le i \le h\).

Wire scans are more complicated than scans of other striation marks in
the literature because:

\begin{enumerate}
\def\labelenumi{\arabic{enumi}.}
\tightlist
\item
  The absence of macro-structure creates alignment issues (i.e.,
  striations are not necessarily vertical but have to be rotated),
\item
  The dome-shaped wire surface introduces structural missing values
  (censored values),
\item
  Vibrations during the scanning process create artificial spikes in the
  observed surface, particularly along the edges of the scanned surface,
\item
  While the wire surface is aligned in parallel to the lens, at the
  micron level, the surface shows strong trends unrelated to the cutting
  procedure,
\item
  Wires tend to roll towards the outer areas of the blades during the
  cutting, resulting in `warped' striation marks.
\end{enumerate}

In order to avoid the subjectivity of manual inspection (and the
resulting variability), we need to address each of these problems in the
algorithmic approach. In the remainder of the paper, we first discuss
the algorithm in Section~\ref{sec-Algorithm} and address each of the
problems listed above. In Section~\ref{sec-Study}, we introduce the
study and its observed data. In Section~\ref{sec-Result}, we show the
results of the algorithm applied to the study's data.

\section{Algorithm}\label{sec-Algorithm}

Figure~\ref{fig-x3ps} shows an overview of the steps from the raw scan
to the extracted signal. We discuss each of the steps in the order of
taking:

\begin{enumerate}
\def\labelenumi{\arabic{enumi}.}
\tightlist
\item
  Identify the boundaries of the wire within the surface matrix \(F\)
  and \textbf{remove edge effects}, Section~\ref{sec-Edgeeffects}.
\item
  Identify and \textbf{remove spikes on the surface},
  Section~\ref{sec-Extremevalues}.
\item
  \textbf{Flatten the surface} by removing trends in the surface
  measurements introduced by scanning positioning,
  Section~\ref{sec-Trend}.
\item
  \textbf{Impute internal missing values} on the surface,
  Section~\ref{sec-Imputation}.
\item
  Identify the main striation direction and \textbf{rotate striations
  into a vertical position}, Section~\ref{sec-Rotation}.
\item
  \textbf{Address warping of striations} introduced by the wire rolling
  during the cutting, Section~\ref{sec-Shifting}.
\item
  Finally, \textbf{extract signals} from the scan by averaging the
  functions \(\tilde{f}_i\) of the processed scan surface \(\tilde{F}\).
\end{enumerate}

The performance of the algorithm is then evaluated on scans where we
know the ground truth, i.e., we know the exact location and blade
involved in those wire cuts. We will measure the similarities between
pairs of scans using the cross-correlation function (CCF) of their
corresponding signals.

For visualization purposes, we discuss the algorithm's steps in the
example of scan \texttt{T1AW-LI-R1} from the data described in
Section~\ref{sec-Study} (see Figure~\ref{fig-x3p}). The naming scheme
tells us that this scan refers to the first (R1) cut made by tool 1 at
the jaw (Location Inner). The scan is of the side engraved by edge A.

\begin{figure}

\begin{minipage}{0.23\linewidth}

\centering{

\captionsetup{labelsep=none}\includegraphics{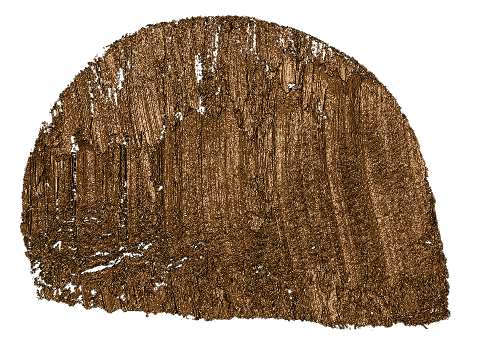}

}

\subcaption{\label{fig-one}}

\end{minipage}%
\begin{minipage}{0.02\linewidth}
~\end{minipage}%
\begin{minipage}{0.23\linewidth}

\centering{

\captionsetup{labelsep=none}\includegraphics{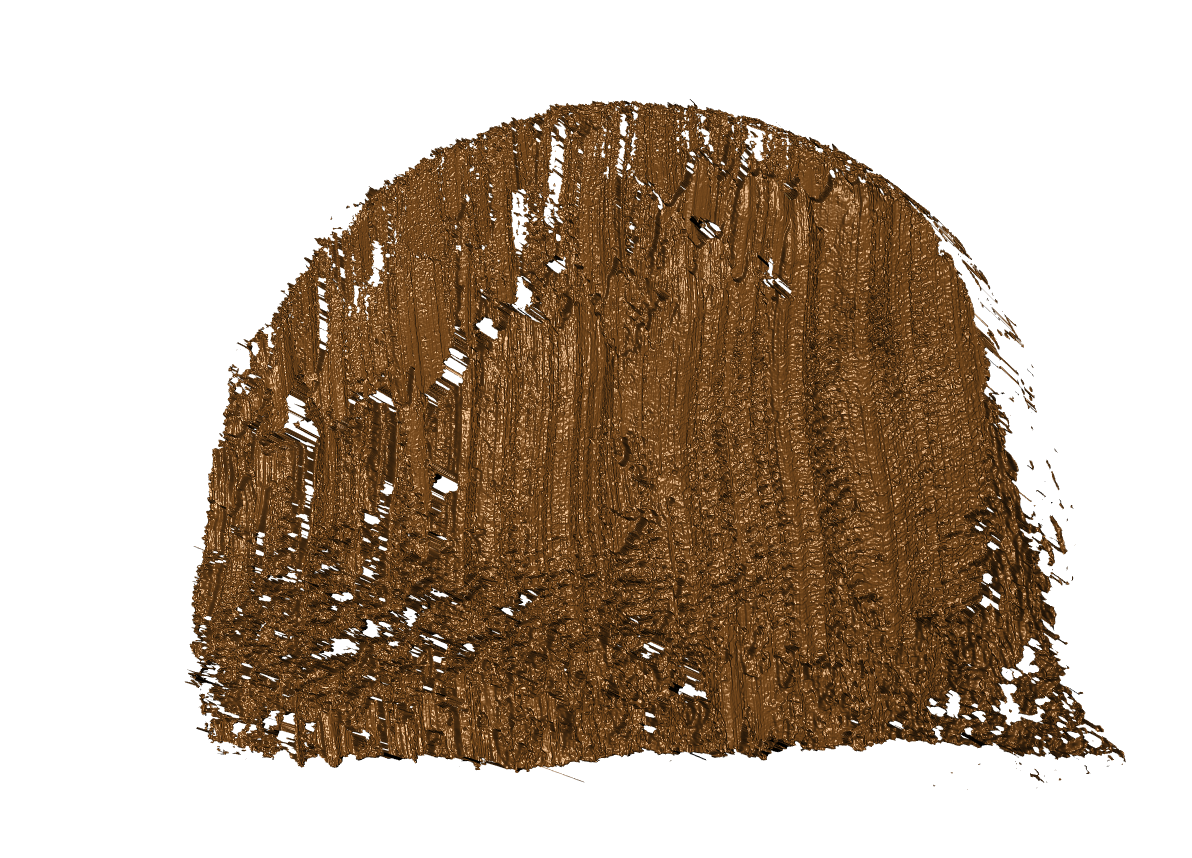}

}

\subcaption{\label{fig-second}}

\end{minipage}%
\begin{minipage}{0.02\linewidth}
~\end{minipage}%
\begin{minipage}{0.23\linewidth}

\centering{

\captionsetup{labelsep=none}\includegraphics{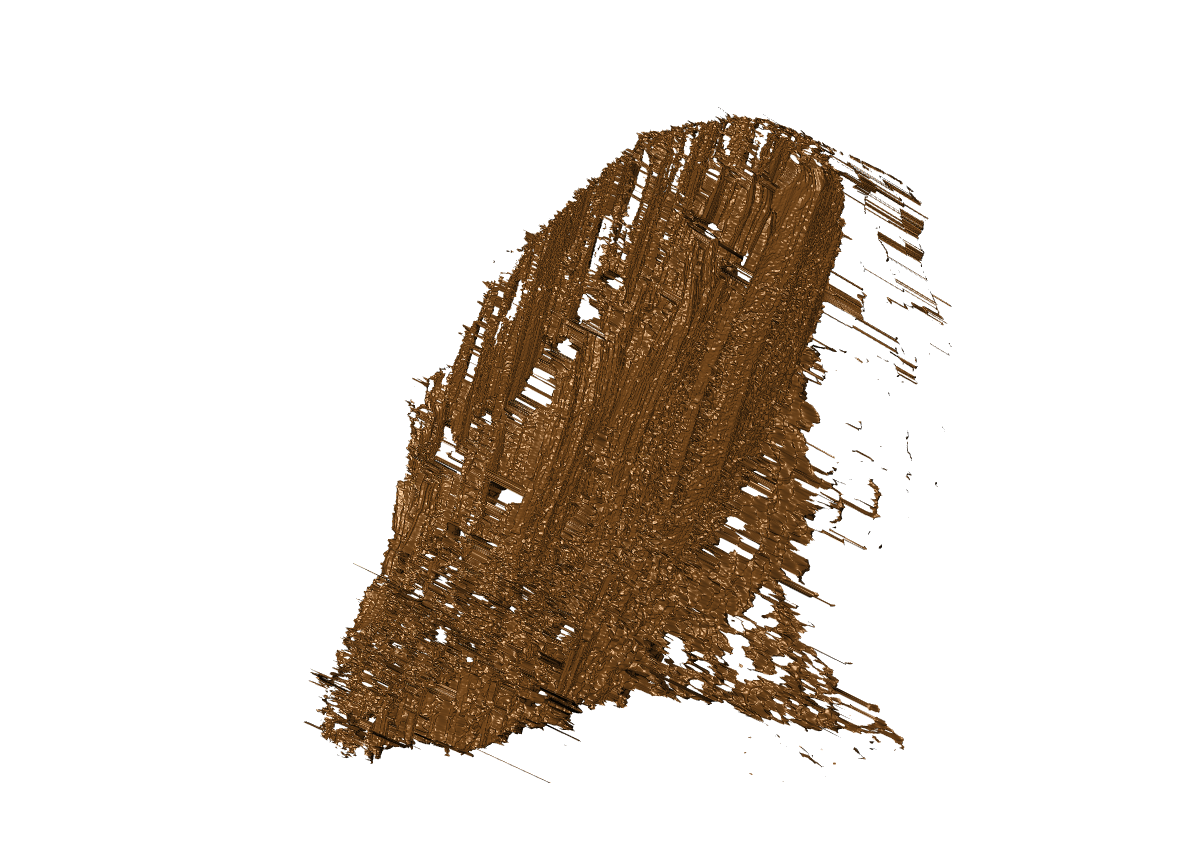}

}

\subcaption{\label{fig-third}}

\end{minipage}%
\begin{minipage}{0.02\linewidth}
~\end{minipage}%
\begin{minipage}{0.23\linewidth}

\centering{

\captionsetup{labelsep=none}\includegraphics{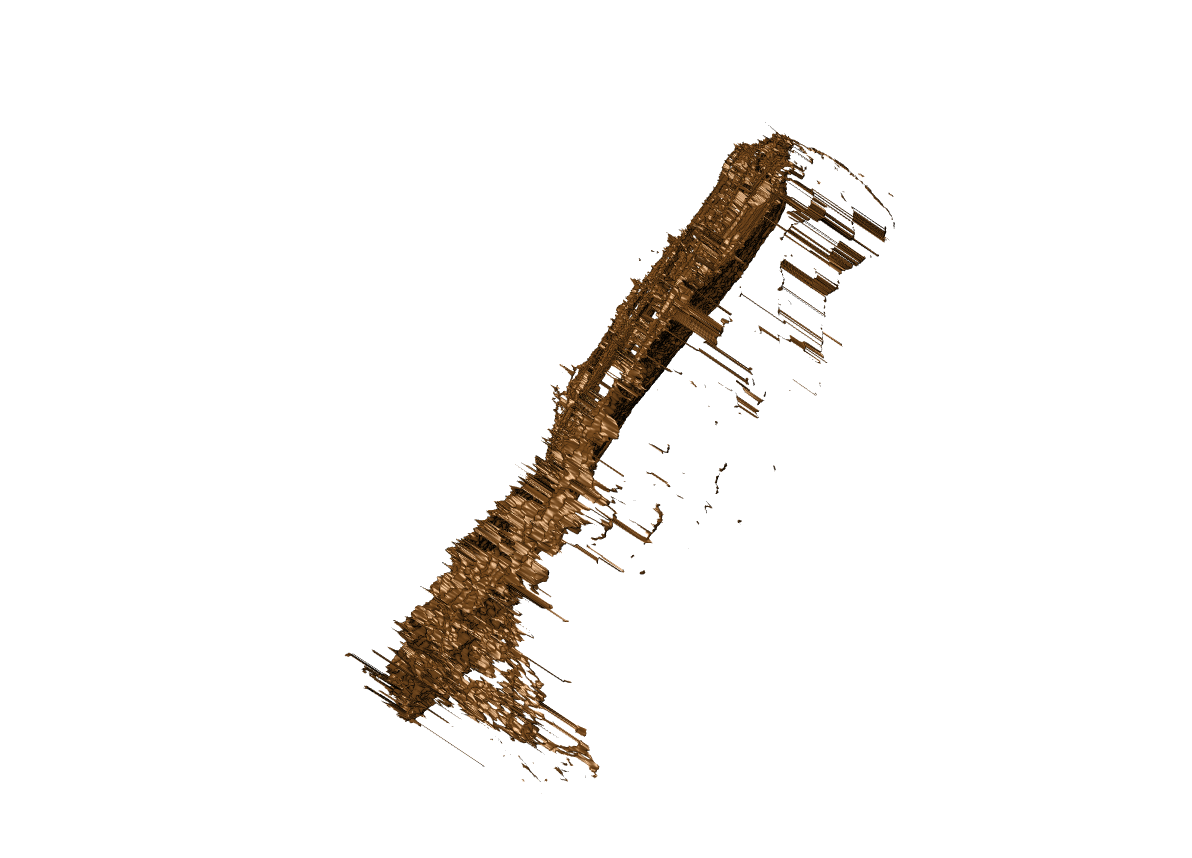}

}

\subcaption{\label{fig-fourth}}

\end{minipage}%

\caption{\label{fig-edges}Tilting the object from the top view (left) to
its side, reveals more and more surface spikes in the surface
measurements. Values along the edges of the scan (right) are
particularly prone to these spikes.}

\end{figure}

The surface matrix \(F\) consists of a regular grid of size
\(2385 \times 1717\) taken at a resolution of
\(0.645\ \mu m \times 0.645\ \mu m\). Figure~\ref{fig-edges} shows that
this scan has large surface spikes along the edges of the wire surface.
We will address these first.

\subsection{Removing edge effects}\label{sec-Edgeeffects}

High-resolution scanning instruments, such as the confocal light
microscope used for the scans in this study, are susceptible to minute
changes. Air perturbations caused by breathing and computer fans cause
the wire to swing (at a sub-micron level), resulting in large spikes
along the edges.

\begin{figure}

\centering{

\includegraphics[width=0.475\textwidth,height=\textheight]{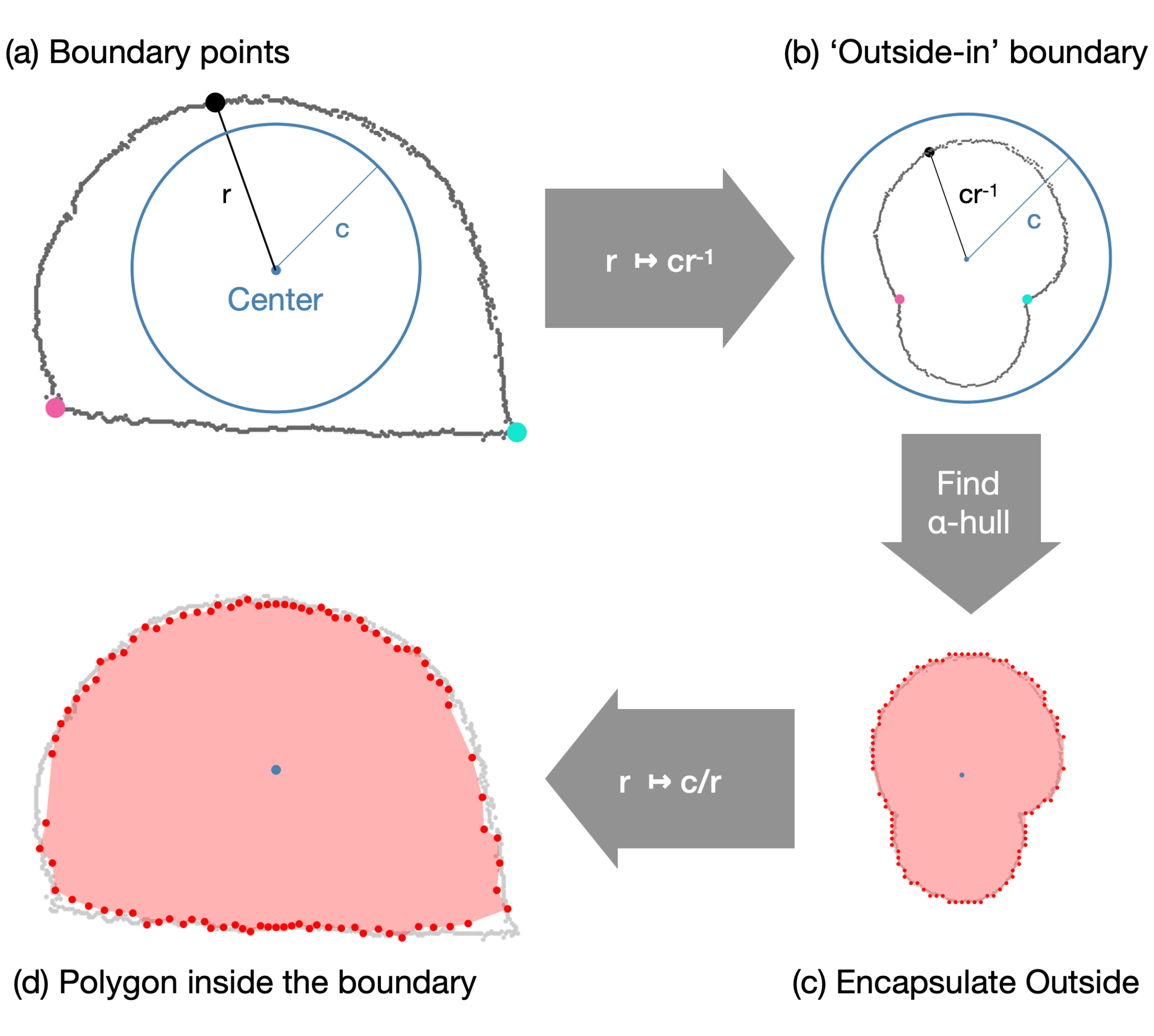}

}

\caption{\label{fig-find-inside-polygon}Process of identifying an inside
mask for a wire cut of arbitrary shape.}

\end{figure}%

\begin{figure*}

\begin{minipage}{0.23\linewidth}

\centering{

\includegraphics{./images/x3p.png}

}

\subcaption{\label{fig-x3p}An overview plot of a sample scan stored in
an x3p object.}

\end{minipage}%
\begin{minipage}{0.02\linewidth}
~\end{minipage}%
\begin{minipage}{0.23\linewidth}

\centering{

\includegraphics{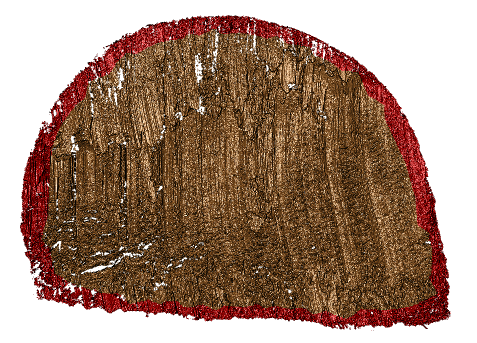}

}

\subcaption{\label{fig-x3p-surface-polygon}Mask added as the concave
polygon for removing edge effects. Only the inner area will be used for
further analysis.}

\end{minipage}%
\begin{minipage}{0.02\linewidth}
~\end{minipage}%
\begin{minipage}{0.23\linewidth}

\centering{

\includegraphics{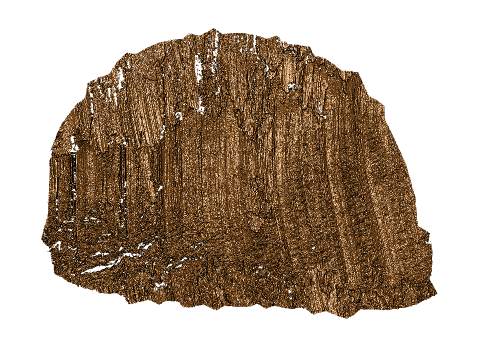}

}

\subcaption{\label{fig-x3p-inner}Concave polygon area used for further
analysis.}

\end{minipage}%
\begin{minipage}{0.02\linewidth}
~\end{minipage}%
\begin{minipage}{0.23\linewidth}

\centering{

\includegraphics{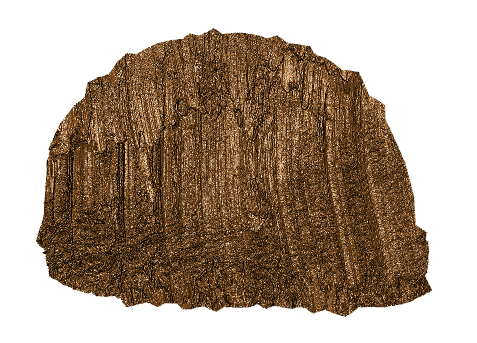}

}

\subcaption{\label{fig-x3p-inner-impute}The surface after imputing all
missing values.}

\end{minipage}%
\newline
\begin{minipage}{0.23\linewidth}

\centering{

\includegraphics{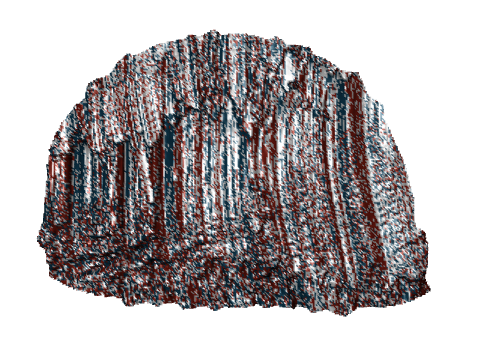}

}

\subcaption{\label{fig-x3p-bin}The surface before rotation with a
colored mask by the values of the sequential differences computed.}

\end{minipage}%
\begin{minipage}{0.02\linewidth}
~\end{minipage}%
\begin{minipage}{0.23\linewidth}

\centering{

\includegraphics{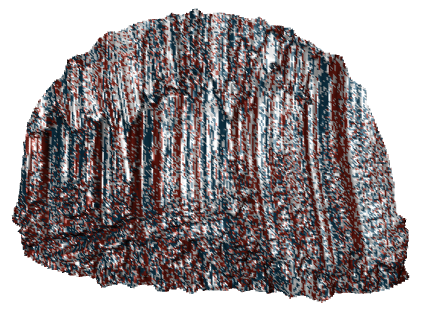}

}

\subcaption{\label{fig-x3p-bin-rotate}The surface after rotation with a
colored mask by the values of the sequential differences computed.}

\end{minipage}%
\begin{minipage}{0.02\linewidth}
~\end{minipage}%
\begin{minipage}{0.23\linewidth}

\centering{

\includegraphics{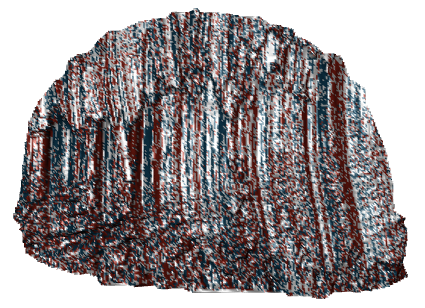}

}

\subcaption{\label{fig-x3p-approx}The surface after shifting striations
by minimizing MSE.}

\end{minipage}%
\begin{minipage}{0.02\linewidth}
~\end{minipage}%
\begin{minipage}{0.23\linewidth}

\centering{

\includegraphics{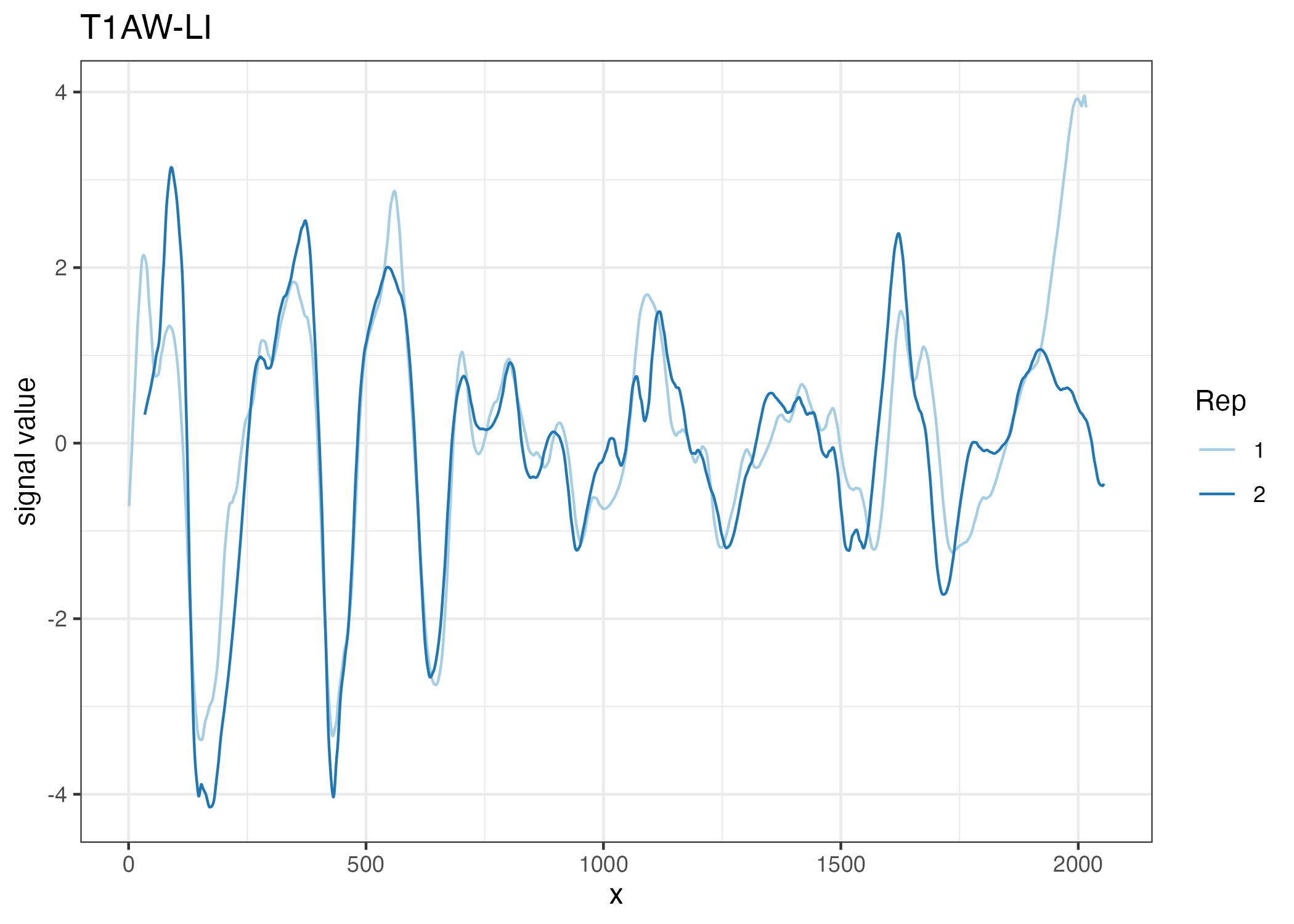}

}

\subcaption{\label{fig-x3p-signals}The extracted signal aligned to the
signal of the repeated cut.}

\end{minipage}%

\caption{\label{fig-x3ps}Overview of the data processing pipeline from
the original scan (top left) to the extracted signal (bottom right).}

\end{figure*}%

These spikes are an order of magnitude bigger than the signal and easily
overwhelm the signal if left unattended. To resolve this problem, we
identify the edge of the scan and remove the points along the
boundary.\\
For that, we find a concave hull in the scan as shown in the sketch in
Figure~\ref{fig-find-inside-polygon}: We first fold all boundary points
to the inside of a circle by applying the function
\(r \mapsto rc^{-1}\), where \(r > 0\) is the distance of a boundary
point from the center of a circle with radius \(c\), \(c > 0\). This
action topologically turns the boundary inside-out, making the problem
of finding a concave hull one of finding a convex hull. We find an
\(\alpha\)-hull of this shape by using the \texttt{concaveman} algorithm
implemented by \citet{gombinConcavemanVeryFast2020}. Folding the
resulting \(\alpha\)-hull back using the same circle center as before
leaves us with a polygon inside the boundaries of the original scan. The
identified boundary is shown in red in
Figure~\ref{fig-x3p-surface-polygon}, and the corresponding points are
removed from the surface matrix (by giving them \texttt{NA} values).

\subsection{Surface spikes}\label{sec-Extremevalues}

\begin{figure}

\centering{

\includegraphics[width=0.45\textwidth,height=\textheight]{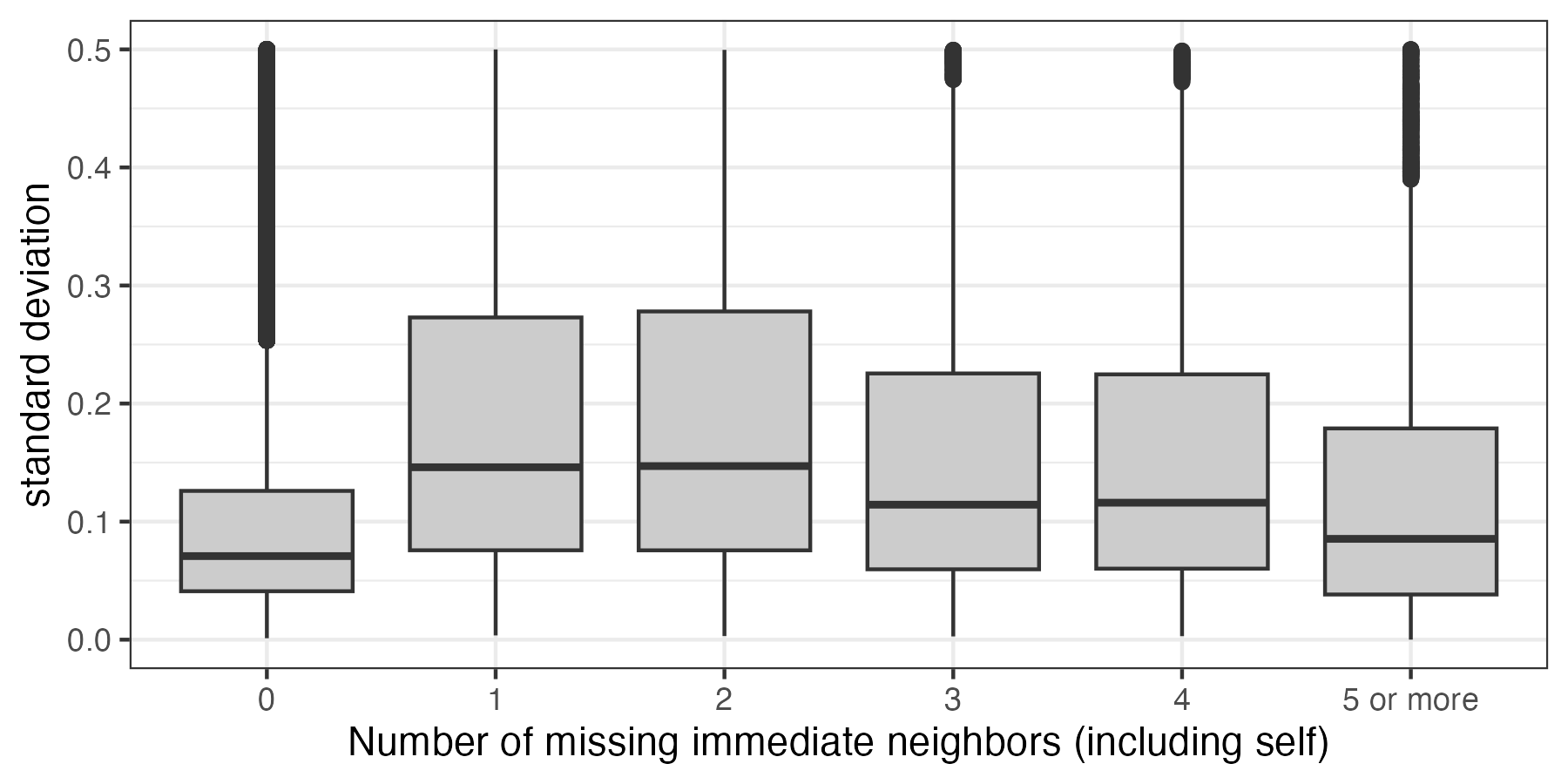}

}

\caption{\label{fig-insidepoly-df-boxplot}Boxplot of standard deviation
against the number of missing immediate neighbors (including self).}

\end{figure}

White areas inside the scan shown in Figure~\ref{fig-x3p} indicate
missing values. These values are not reported by the microscope because
their exact value cannot be determined accurately enough.
Figure~\ref{fig-insidepoly-df-boxplot} shows boxplots of the standard
deviation observed conditioned on the number of missing values in the
immediate neighborhood. Note that the distribution of standard
deviations is extremely skewed. Fewer than 75\% of all values are larger
than 0.5, but the overall range of standard deviations in this example
reaches a value of 20 microns. We see from the figure that values around
dropouts show an inflated variability -- even just one missing value
among the immediate neighbors increases the median of the standard
deviation by a factor of about 3.

So, to eliminate these extreme values, we only keep those points without
any missing values in the neighbor. While this might seem like an
extreme measure, it only affects a small percentage of observations. In
the scan of Figure~\ref{fig-edges}, 2.7\% (67518 measurements) of the
observations are thus affected.

\subsection{Trend}\label{sec-Trend}

Now that the most extreme values are removed from the surface scan, we
can consider the structure of the signal we want to extract. We want to
extract a representative curve as our signal, formed by the ups and
downs of the striation. If we extract the signal directly from the
surface as it is now, the curvature of the surface dominates the
behavior of the signal. Therefore, we need to remove the trends before
proceeding. For that, we use regression with quadratic terms and an
interaction effect between rows and columns of the surface matrix:
\begin{align*}
    F_{ij} = \beta_0 + \beta_1 x_i + \beta_2 x_i ^ 2 + \beta_3 y_j + \beta_4 y_j ^ 2 + \beta_5 x_i y_j + \epsilon_{ij},
\end{align*} Note that the coefficients
\(\beta_0, \beta_1, \beta_2, \beta_3, \beta_4, \beta_5\) are nuisance
parameters, i.e., the whole fit is only used to remove any trend in the
surface matrix. Rather than using \(\hat{F}_{ij}\) of the above fit, we
modify the surface values by correcting for this fit. The new values of
\(\tilde{F}_{ij} = \epsilon_{ij}\) reflect the surface values that are
unaffected by the trend and capture the signal engraved by the tool
rather than the positioning of the wire under the microscope.

\subsection{Imputation}\label{sec-Imputation}

After separating the overall trend and the noise, we focus now on
missing values on the surface, either existing from the beginning or
created by the algorithm when removing spikes in
Section~\ref{sec-Extremevalues}. We fill these holes by iteratively
imputing missing values as an average of values of the immediate
neighborhood of a \(3 \times 3\) grid, i.e., using the same immediate
neighbors as before, until we get a full surface without any missing in
the original scan area. We use the boundary shape from
Figure~\ref{fig-x3p-inner} to restrict the surface to interpolated
values. Points outside the boundary result from extrapolation, which can
be highly biased and extreme. The surface after imputation is plotted in
Figure~\ref{fig-x3p-inner-impute}.

\subsection{Rotation}\label{sec-Rotation}

\begin{figure}

\begin{minipage}{0.48\linewidth}

\centering{

\includegraphics{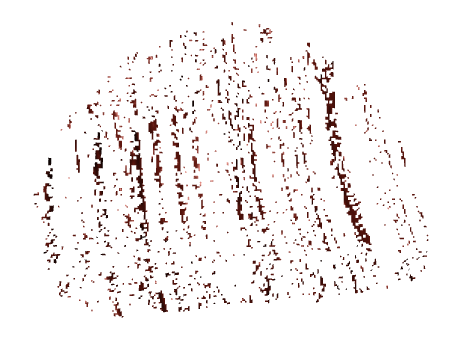}

}

\subcaption{\label{fig-x3p-bin-red}Areas of steepest decline.}

\end{minipage}%
\begin{minipage}{0.04\linewidth}
~\end{minipage}%
\begin{minipage}{0.48\linewidth}

\centering{

\includegraphics{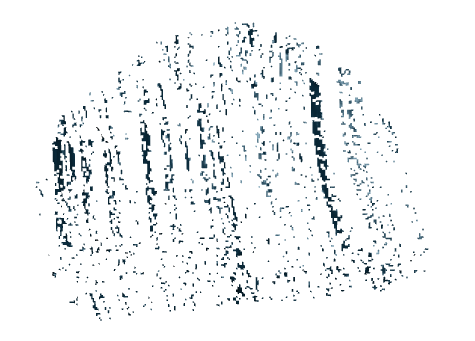}

}

\subcaption{\label{fig-x3p-bin-blue}Areas of steepest incline.}

\end{minipage}%

\caption{\label{fig-x3p-bins}Regions of steepest decline (red) and
incline (blue).}

\end{figure}

In order to align striations vertically, we have to first identify
striations -- a stria is defined as a valley engraved by a
micro-imperfection `sticking out' from the surface of a tool's blade. We
can identify the sides of these valleys in each row of the surface
matrix \(F\) by using a lag 1 difference
\(DF_{i,j} = F_{i, j+1} - F_{i, j}\) for \(1 \le i \le h\) and
\(1 \le j < w\).\\
Figure~\ref{fig-x3p-bins} shows areas of the steepest decline (left) and
incline (right).

The stripes we see in these images follow the slopes of the striation
marks and, therefore, serve as a good representation of the angle under
which the striation marks are located in the scan.

In Figure~\ref{fig-x3p-bin}, the images of Figure~\ref{fig-x3p-bins} are
overlaid as masks. Note that in order to better visualize the striation
marks, we down-sampled the scans shown in Figure~\ref{fig-x3p-bin} to
Figure~\ref{fig-x3p-approx} considerably (by a factor of 8) before
plotting.

\begin{figure}

\centering{

\includegraphics[width=0.45\textwidth,height=\textheight]{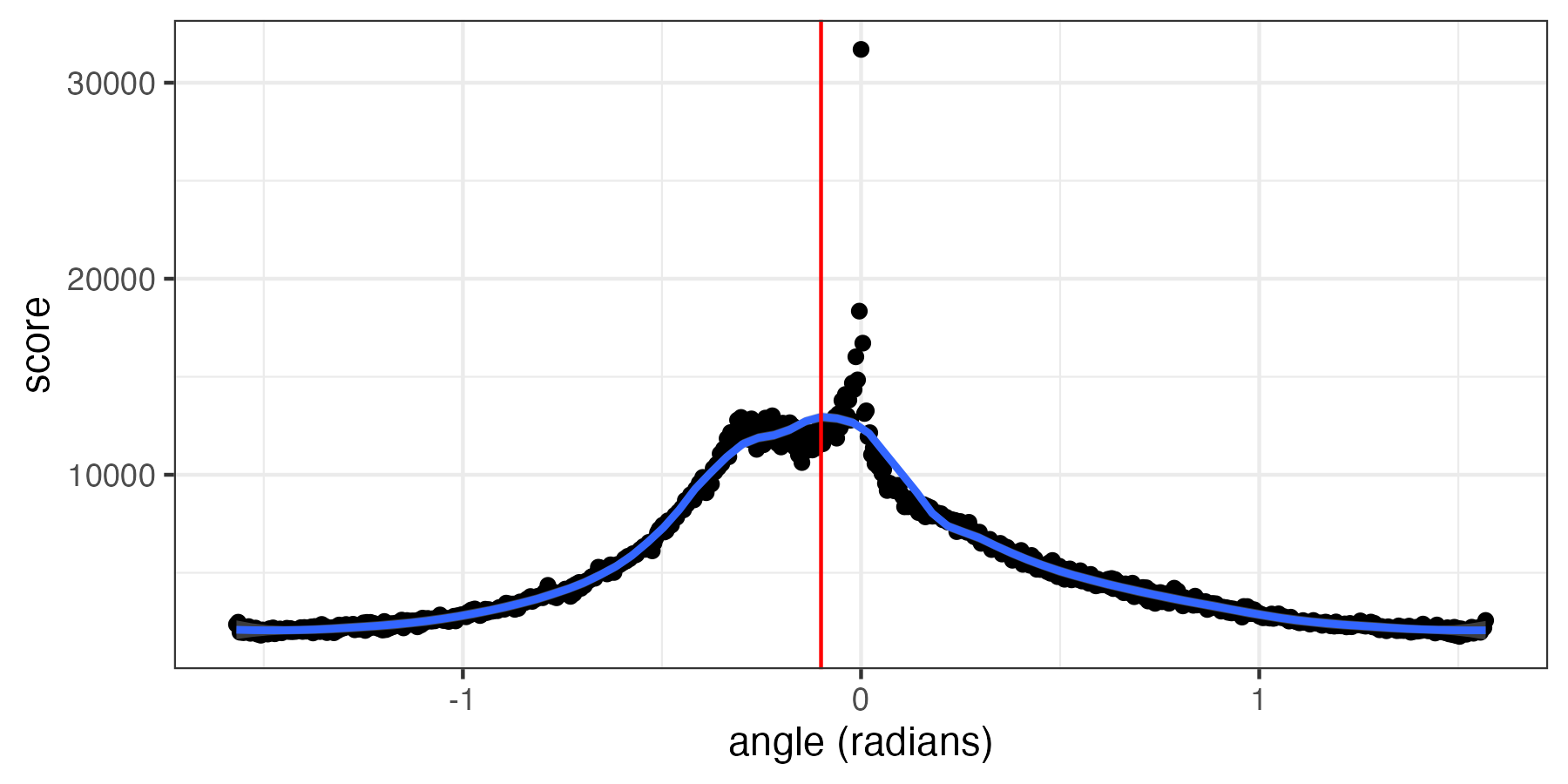}

}

\caption{\label{fig-x3p-hough-theta-loess}Loess fit for angle in
radians.}

\end{figure}

\begin{figure}

\centering{

\includegraphics[width=0.4\textwidth,height=\textheight]{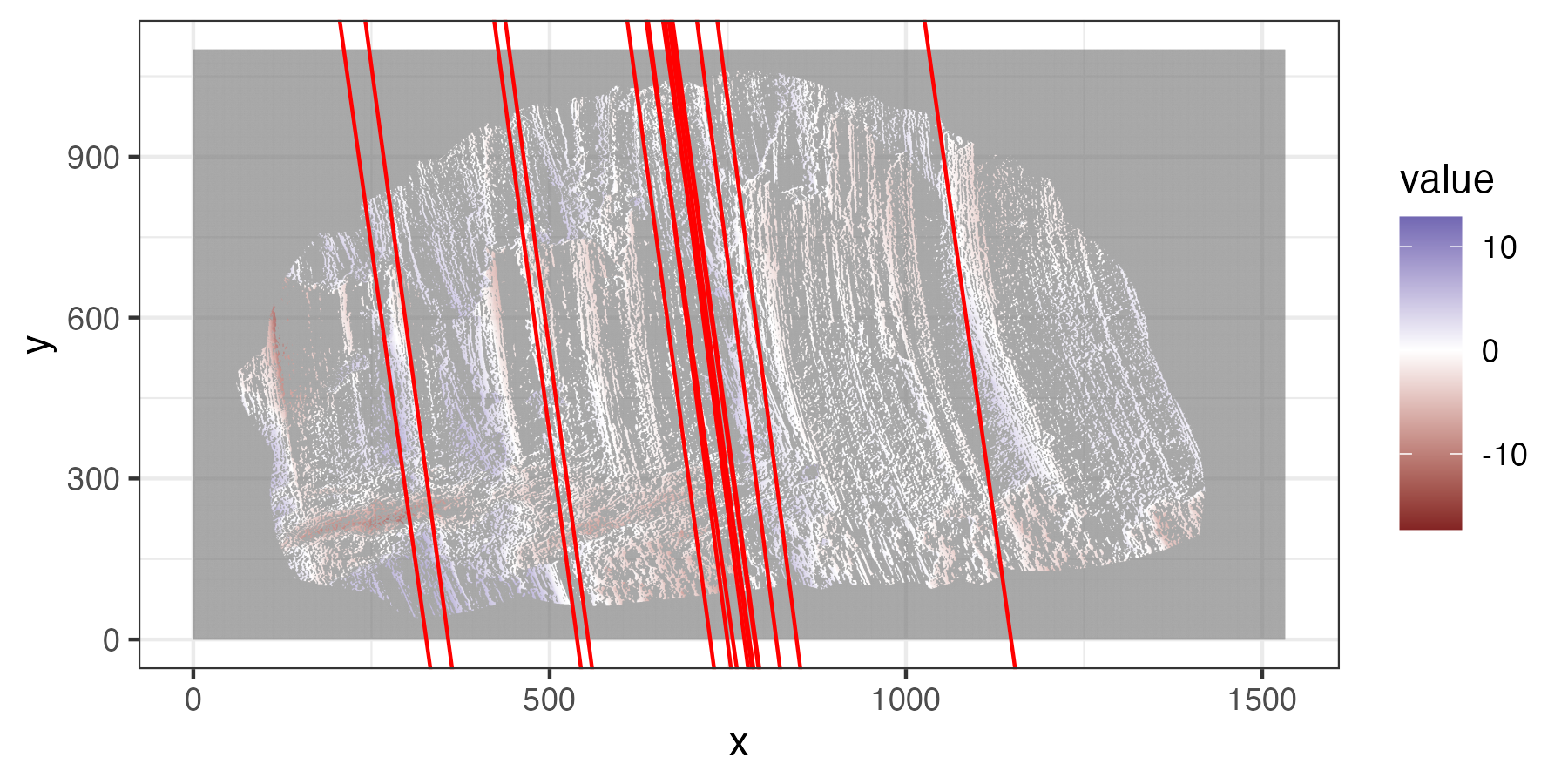}

}

\caption{\label{fig-x3p-cimg-nfline}Hough lines detected before rotation
are marked as red.}

\end{figure}

We use a Hough transformation \citep{dudaUseHoughTransformation1972} to
obtain pairs of polar coordinates \((r, \theta)\) that represent the
directions of possible vertical lines on the picture, also known as
Hough lines. Then, our goal is to estimate the main direction \(\theta\)
of the lines. We numerically integrated over radius \(r\) and found the
distribution of these angles, shown in
Figure~\ref{fig-x3p-hough-theta-loess}. The red vertical line indicates
the maximum likelihood estimate (MLE) for \(\theta\), which we derive
based on using a LOESS (locally estimated scatterplot smoothing) of the
density as shown by the blue line in
Figure~\ref{fig-x3p-hough-theta-loess}, with the red line showing the
MLE. The Hough lines corresponding to the estimated value of
\(\hat{\theta}\) are shown as red overlays in
Figure~\ref{fig-x3p-cimg-nfline}. Rotating the scan by \(\hat{theta}\)
turns most of the striations into a vertical position along the the
scan. The resulting scan is shown in Figure~\ref{fig-x3p-bin-rotate}.

\subsection{Shifting}\label{sec-Shifting}

During the cutting process, the wire rolls (at a microscopic level) away
from the tool's jaw, resulting in curved striation marks. This curvature
is visible in the scans of the previous section. The density of the
Hough line angles in Figure~\ref{fig-x3p-hough-theta-loess} also
suggests a secondary mode with an angle more extreme than the identified
\(\hat{\theta}\).

We deal with this curvature by using small, line-wise horizontal shifts
of the signals by minimizing the mean square error (MSE) for all \(f_i\)
with respect to a base signal \(f_0\): \begin{align*}
  \text{MSE}_{i k} =
  \frac{1}{n_{i k}} \sum_{j = 1} ^ {n_{i k}} \left(f_{0 (j+k)} - f_{i (j+k)} \right) ^ 2
  ,
\end{align*} where \(n_{i k}\) is the count for the number of
non-missing values for \(f_i\), \(f_{0 (j+k)}\) is the \((j + k)\)th
element of \(f_0\), \(f_{i (j+k)}\) is the \((j + k)\)th element of
\(f_i\), where \(-\delta \le k \le \delta\).

\begin{figure}

\centering{

\includegraphics[width=0.45\textwidth,height=\textheight]{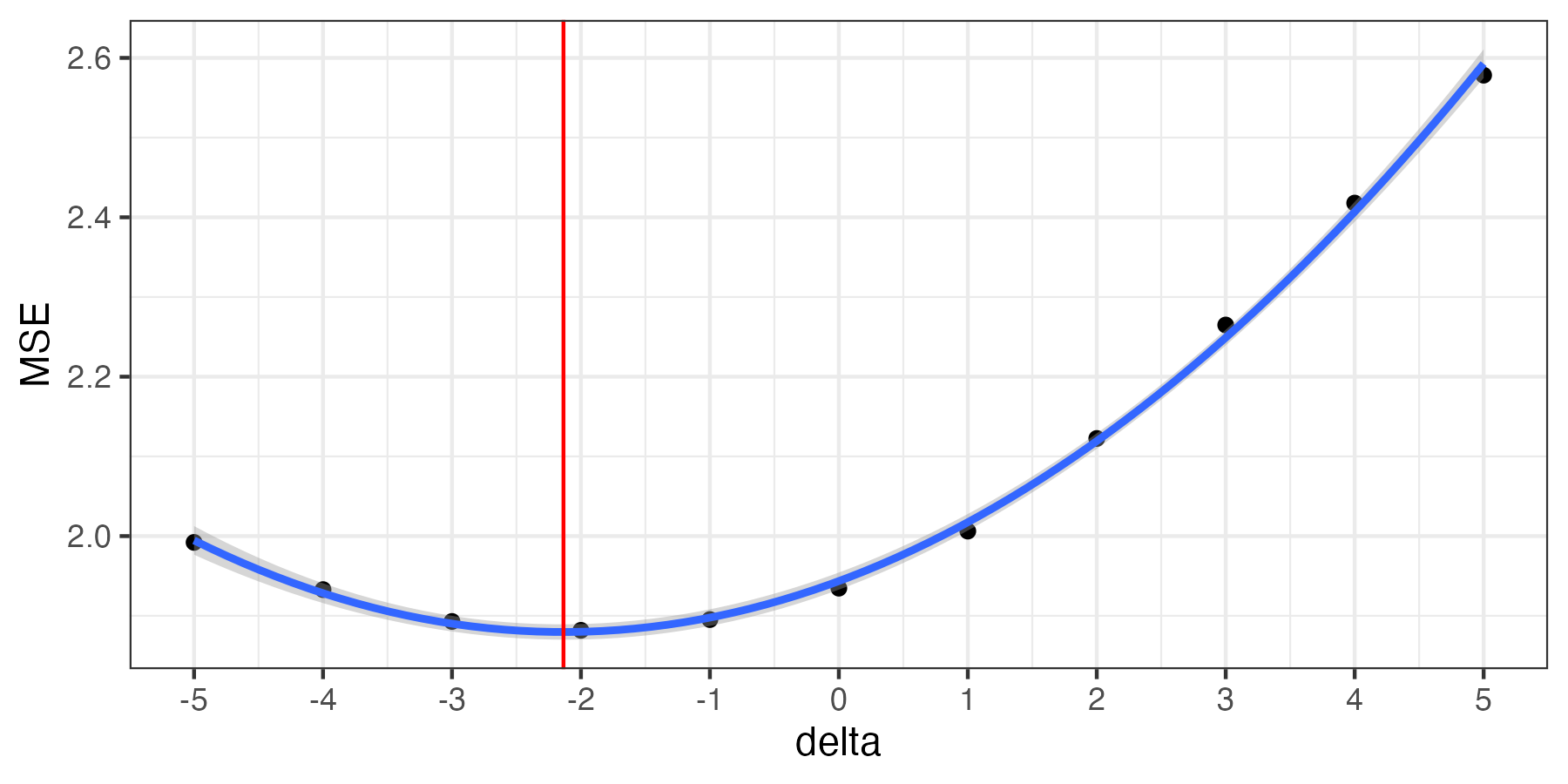}

}

\caption{\label{fig-x3p-approx-MSE}Parabola fitted with shifting value
with minimum MSE marked as red.}

\end{figure}

While we evaluate the above function only for integer values of \(k\),
we fit a parabola for each pair and obtain the desired shift width that
minimizes the MSE, which can be a non-integer value bounded in the given
range, as shown in Figure~\ref{fig-x3p-approx-MSE}. These shifting
widths then move curves for each \(y\) along the \(x\) direction. The
resulting surface is shown in Figure~\ref{fig-x3p-approx}.

\section{Results}\label{sec-Result}

\subsection{Study}\label{sec-Study}

In our study, we will focus on wire cutters.

To begin the study, we first need data, including some from the crime
scene and others from a given tool. In the development process of the
whole pipeline, we should have all marks created by known tools, which
will be regarded as the ground truth when comparing either pair of marks
and computing the error rates. So, we prepared 5 Kaiweets wire cutters
(model KWS-105) and aluminum wire (16 Gauge/1.5 mm, anodized). We
labeled all sides of the blades of wire cutters as AB CD, shown in
Figure~\ref{fig-overview} a, and used them to cut at 3 different
locations, inner, middle, and outer, along the blade twice. As a result,
we have \(5 \times 2 \times 3 \times 2 = 60\). The cutting surface for
each side of AB and CD formed tent structures as in
Figure~\ref{fig-overview} c that can be separated into 2 parts, long
edge A and short edge B, long edge C and short edge D. So, we scan each
edge by a confocal light microscope as in Figure~\ref{fig-overview} b,
resulting in \(60 \times 2 = 120\) marks in total.

\subsection{Data structure}\label{sec-Datastructure}

These scans are stored as \texttt{x3p} objects, a file format specified
in ISO standard 25178-72:2017/AMD 1:2020 (based on ISO ISO5436 -- 2000)
describing 3D surface measurements at a resolution of
\(0.645 \mu m \times 0.645 \mu m\) per square pixel. The naming scheme
for these files consists of 4 parts: tool (\texttt{T1}, \texttt{T2},
\texttt{T3}, \texttt{T4}), edge (\texttt{A}, \texttt{B}, \texttt{C},
\texttt{D}), location (\texttt{I}, \texttt{M}, \texttt{O}), and
repetition (\texttt{R1}, \texttt{R2}). For example, a file named
\texttt{T1AW-LI-R1} is the first replicate made using the inner location
of edge A of tool 1. All these 120 well-documented \texttt{x3p} objects
will be uploaded to a public data repository, and a more detailed data
description paper can be found here.

With this set of data in position, given any pairs, our question is, can
we say both marks are made from the same tool, and how accurate is our
conclusion? Because both marks are made from known tools 1 to 5, we can
compute the error rate of our algorithm by matching our conclusion with
the truth. Later, we can apply the same algorithm to data from real
crime scenes and get the decision.

\subsection{Evaluation}\label{evaluation}

So, after all previous efforts, we can now extract signals by computing
the median of values for each \(x\) along the \(y\)-axis. In
Figure~\ref{fig-x3p-signals}, signals extracted from 2 replicates of
\texttt{T1AW-LI} cuts are aligned together.

Then, we can use these extracted signals for alignment and comparison by
maximizing CCF.

\begin{figure}

\centering{

\includegraphics[width=0.45\textwidth,height=\textheight]{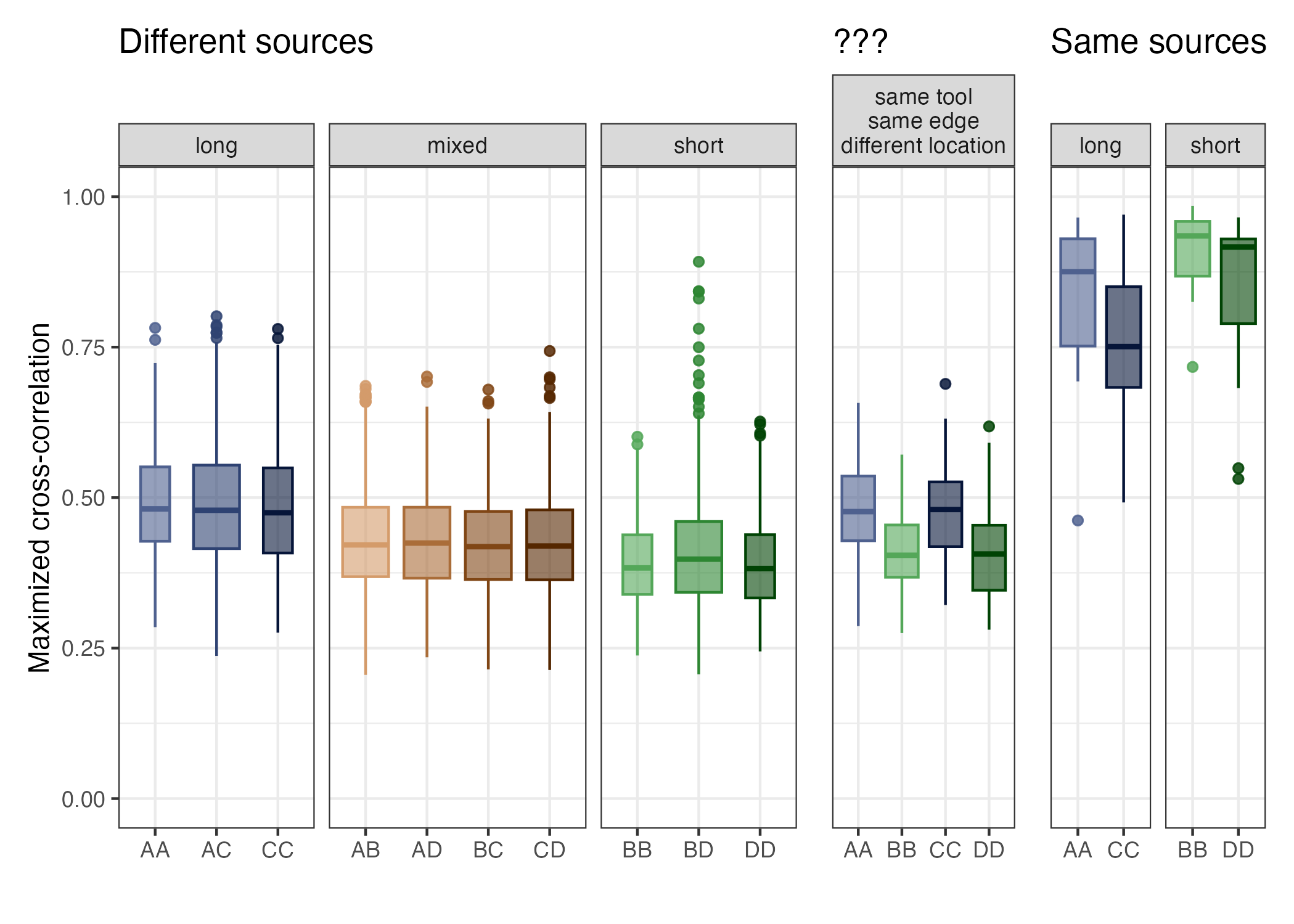}

}

\caption{\label{fig-ccf-boxplot}Boxplots of resulting CCFs for all cut
pair combinations.}

\end{figure}

Assessing all pairwise signals based on their maximized
cross-correlation results in a decent amount of separations between
known same-source and known different-source pairs. All combinations of
\(120 \times 119 = 14,280\) different pairwise signals are considered
here and take less than 2 hours to run without any kind of parallel
computing, which does not cause any computational burden. Boxplots of
resulting CCFs are shown in Figure Figure~\ref{fig-ccf-boxplot}. Here,
pairs with the same sources have an obviously high CCF compared to those
with different tools and edges or locations, marked as \texttt{???} on
the plot.

\begin{figure}

\centering{

\includegraphics[width=0.45\textwidth,height=\textheight]{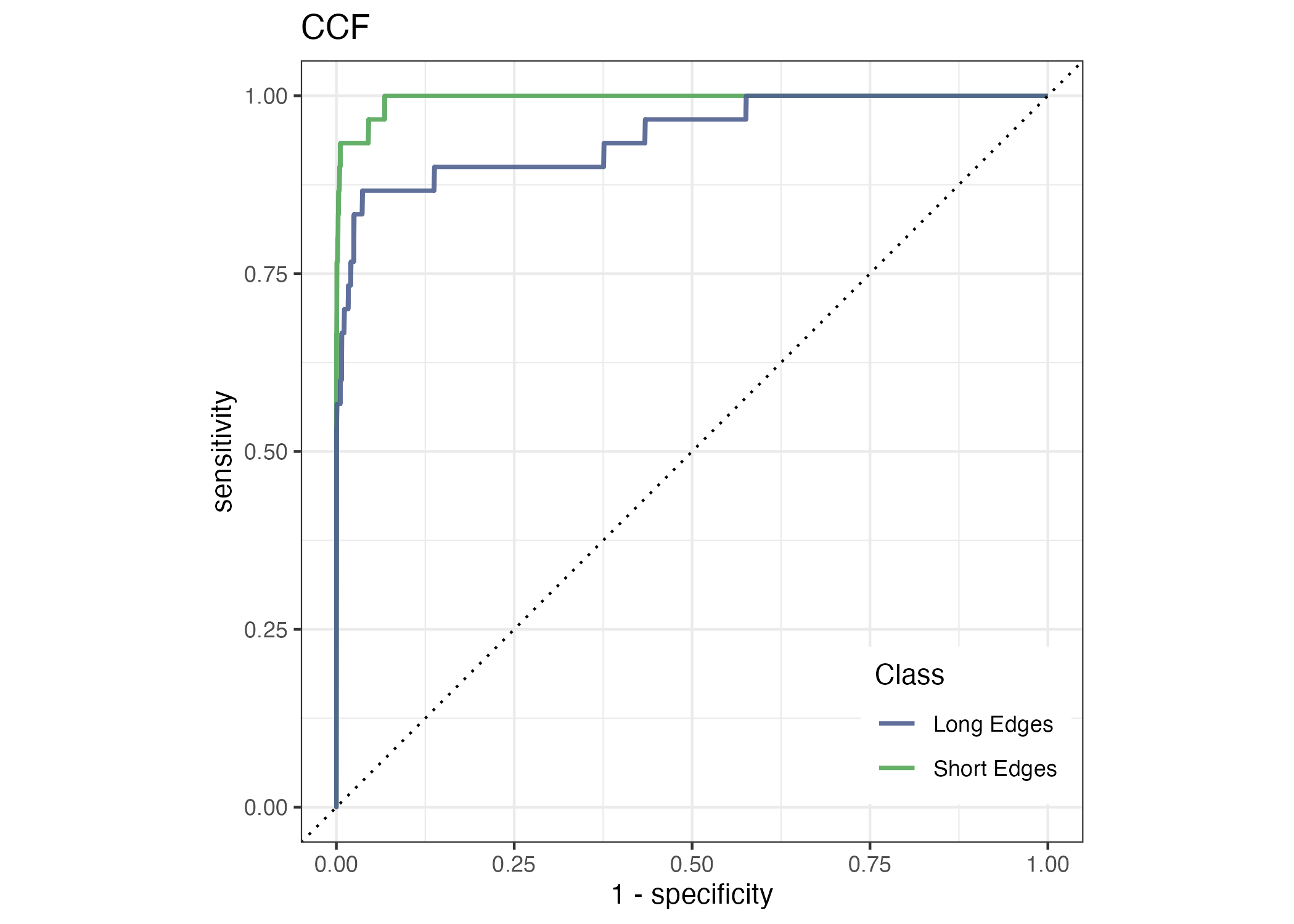}

}

\caption{\label{fig-ccf-ROC}ROC curve for different CCF threshold.}

\end{figure}

We also have the receiver operating characteristic (ROC) curve for
different CCF threshold plotting in Figure~\ref{fig-ccf-ROC}. The ROC
curve tells us the classifier used in our algorithm is good as it goes
toward the upper left corner \((0, 1)\). We can also know that for a CCF
threshold of 0.68, an FPR of less than 5\% is achieved, with an FNR of
about 23\%. Increasing the CCF threshold to 0.78 reduces the FPR to
below 1\%, while the FNR increases to about 40\%. These results are
comparable to other algorithmic similarity assessments of striation
marks.

Last but not least, all features discussed before are wrapped and
documented into functions with reproducible minor examples in our
developmental \texttt{R} package \texttt{wire} on GitHub, which will be
submitted to \texttt{CRAN} afterward.

\renewcommand\refname{References}
  \bibliography{bibliography.bib}

\end{document}